# The Optimal Geometry of the Acousto-Optic Interaction in Selected Crystalline Materials Determined by Extreme Surfaces Method

Anatoliy Andrushchak
*Telecommunications Department*
*Lviv Polytechnic National University*
Lviv, Ukraine
anatolii.s.andrushchak@lpnu.ua

Oleh Buryy
*Semiconductor Electronics Department*
*Lviv Polytechnic National University*
Lviv, Ukraine
oleh.a.buryi@lpnu.ua

*Abstract*— The optimal geometries of the acousto-optic interaction are determined by extreme surfaces method for a number of acousto-optic crystals.

## I. INTRODUCTION

The acousto-optic (AO) effect i.e. the diffraction of light on the irregularities of dielectric permittivity caused by acoustic wave propagation in the transparent medium, has numerous uses in science and technology [1–3]. The most efficient operation of AO devices – modulators, deflectors, filters, etc. can be achieved by the determination of such directions of light and acoustic waves that maximize the value of the effect (i.e. by finding of the optimal AO interaction geometry). This optimization can be carried out by extreme surfaces method elaborated in [4–7]. Here we present the results of this optimization for a number of crystalline materials widely used in acousto-optics or promising for it. At that only the Bragg diffraction regime is considered because of its prevalence in AO devices.

## II. METHOD OF ANALYSIS

As it is known, the efficiency of Bragg diffraction $\eta_{ef}$ at a given acoustic wave intensity $I_s$ is determined by the expression [3]:

$$\eta_{ef} = \frac{I_\nu}{I_\mu} = \sin^2\left(\frac{\pi L}{\sqrt{2}\lambda}\sqrt{M_2 I_s}\right). \qquad (1)$$

Here $I_\mu$ and $I_\nu$ are the intensities of the incident and diffracted electromagnetic waves correspondingly, $L$ is the length of the region of AO interaction, $\lambda$ is the wavelength of light, $M_2$ is the AO figure-of-merit that is in accordance with the consideration by Nelson [8,9] is equal to,

$$M_2 = \frac{n_\mu^3 n_\nu^3}{\rho V_q^3} p_{ef}^2 \cos\gamma_\nu, \qquad (2)$$

where $n_\mu$ and $n_\nu$ are the refractive indices of the incident and diffracted electromagnetic waves correspondingly, $\rho$ is the density, $V_q$ is the acoustic wave velocity, $\gamma$ is the drift angle of the acoustic wave, $p_{ef}$ is the effective elasto-optic coefficient,

$$p_{ef} = \vec{i}_\mu \vec{i}_\nu \hat{p} \vec{a} \vec{f}_q, \qquad (3)$$

$\vec{i}_\mu$, $\vec{i}_\nu$ and $\vec{f}_q$ are the unit vectors determining the directions of polarizations of the incident, diffracted and acoustic waves correspondingly, $\vec{a}$ is the normal vector of the acoustic wave, $\hat{p}$ is the tensor of elasto-optic coefficients. The velocities $V_q$ and the polarizations $\vec{f}_q$ of the acoustic waves for each direction of $\vec{a}$ can be determined from Christoffel' equation [10]:

$$\left[\hat{\vec{a}}\hat{c}\vec{a} + \frac{\left(\hat{\vec{a}}\hat{e}\vec{a}\right)\left(\hat{\vec{a}}\hat{e}\vec{a}\right)}{\varepsilon_0\left(\hat{\vec{a}}\hat{\varepsilon}\vec{a}\right)}\right]\vec{f}_q = \rho V_q^2 \vec{f}_q. \qquad (4)$$

Here $\hat{c}$, $\hat{e}$ and $\hat{\varepsilon}$ are the tensors of elasticity, piezoelectricity and dielectric permittivity (at the frequency of the acoustic wave) correspondingly. The expression $\left(\hat{\vec{a}}\hat{e}\vec{a}\right)\left(\hat{\vec{a}}\hat{e}\vec{a}\right)$ denotes a dyadic tensor obtained as a tensor product of vectors $\left(\hat{\vec{a}}\hat{e}\vec{a}\right)$.

The optimization is carried out in the following way. For each incident light direction defined by polar and azimuthal angles $\theta$ and $\varphi$ we determine the corresponding direction of the acoustic wave (defined by angles $\theta_a$, $\varphi_a$) that allows to obtain the maximal value of the AO figure-of-merit $M_2$. At that the direction of the diffracted light wave is determined in accordance with the momentum conservation law [1]

$$\vec{k}_\nu = \vec{k}_\mu \pm \vec{K}, \qquad (5)$$

where $\vec{k}_\nu$, $\vec{k}_\mu$ and $\vec{K}$ are the wave vectors of the diffracted ($\nu$), incident ($\mu$) and acoustic waves correspondingly. It is essential for the optimization procedure, that the relation (5) limits the search area of the optimal values of $\theta_a$ and $\varphi_a$ [4]. The dependencies of the maximal $M_2$ values on angles $\theta$ and $\varphi$ can be presented by special type ('extreme') surfaces. Characterization of AO interaction at the given light wavelength and acoustic frequency requires construction of 12 extreme surfaces because of four possible types of diffraction depending on polarizations of the incident and diffracted light waves and three possible types of acoustic wave: quasi-longitudinal (*l*), fast quasi-transversal (*f*) and slow quasi-transversal (*s*) ones.

Here the optimization is carried out for a number of AO crystals, particularly, LiNbO$_3$:MgO [11], β-BaB$_2$O$_4$ [12], Cs$_2$HgCl$_4$ [13], SiO$_2$ [14], Li$_2$B$_4$O$_7$ [15]. For comparison we also give the results of optimization for LiNbO$_3$, SrB$_4$O$_7$, La$_3$Ga$_5$SiO$_{14}$ (langasite), CaWO$_4$ and GaP crystals investigated in [4–7, 16].

## III. Results and discussion

The examples of the extreme surfaces for the considered crystals are shown in Fig. 1 – 2 (except for LiNbO$_3$ crystal, because the corresponding extreme surface is visually the similar to the one for LiNbO$_3$:MgO). The shown surfaces correspond to the type of diffraction that reveals the highest possible AO figure-of-merit $M_{2surf}^{max}$. In the cases of SiO$_2$ and CaWO$_4$ crystals the same maximal value of $M_{2surf}^{max}$ is achieved for different types of diffraction, so two extreme surfaces are shown for SiO$_2$ and three ones for CaWO$_4$. The wavelength of light beams in all cases is equal to 633 nm.

The evident notation specified the type of diffraction is used in Figs. 1 – 2: the first and the last symbols in the designations indicate the polarizations of the incident and diffracted beams. As usual, in the case of uniaxial crystals '$o$' designates the ordinary light beam, '$e$' – the extraordinary one, whereas for biaxial crystals '$i$' and '$e$' correspond to light waves which wave vectors end on the internal ($i$) or external ($e$) parts of the wave vectors surface (see [5] for details). The symbol over the arrow indicates the type of the acoustic wave ($l$, $f$ or $s$).

As it is seen from Figs. 1 – 2, the symmetry of the extreme surfaces is determined by the point group symmetry of the crystal (indicated in the Figures) complemented by inversion if the crystal is non-centrosymmetrical one, because the changes of directions of wave vectors or polarizations on the opposite ones do not change the sign of $p_{ef}$ and, consequently, $M_2$. The general form of the extreme surfaces is not the same for the crystals of the same point group (e.g. see Fig. 1 where the extreme surfaces for the trigonal crystals are shown) but strongly depends on the values of the elastic and elasto-optic coefficients similar to the cases of linear electro-optic and piezo-optic effects [17,18].

The results of optimization, particularly, the characteristic of the incident and acoustic waves (optimal directions and polarizations, the velocity of the acoustic wave), the corresponding value of the effective elasto-optic coefficient $p_{ef}$ and the highest achievable value of $M_{2surf}^{max}$ are indicated in Table 1. It should be noted that only one of the few symmetrically equivalent directions is given in the Table. The values of the frequency $f$ of the acoustic waves used in our calculations are also mentioned in the Table 1. The choice of the frequency in each case is caused by the possible using of the considered crystals. The highest achievable value of $M_2$ = 115.9·10$^{-15}$s$^3$/kg is obtained for the optically biaxial Cs$_2$HgCl$_4$ crystal, highlighting the perspectiveness of low symmetry crystalline materials using in AO devices. Certainly, the practical choice of the material is based not only on the value of $M_2$, but realized in accordance with a combination of properties. Particularly, the high transmittance of SiO$_2$, SrB$_4$O$_7$ and CaWO$_4$ crystals in the ultra-violet spectral range (down to 130 … 150 nm [6, 14, 19]) allows their using in the near-UV acousto-optic devices (the short wavelength cut-off of Li$_2$B$_4$O$_7$ is also close and is observed at 167 nm [20]). At that the relatively low attenuation of the acoustic waves in CaWO$_4$ crystal is the additional advantage of this material [6]. On the other hand, GaP crystal reveals high value of $M_2$ = 97.6·10$^{-15}$s$^3$/kg, however, its using in the visible region is limited by the wavelength of about 550 nm [21].

As it is seen from Table 1, the global maximum of $M_2$ for the cases of LiNbO$_3$, LiNbO$_3$:MgO, BBO, Cs$_2$HgCl$_4$, SiO$_2$ is achieved on indirect cuts of the crystals. For other crystals the optimal cut is direct (SrB$_4$O$_7$) or close to it (langasite, CaWO$_4$, Li$_2$B$_4$O$_7$). In the case of GaP crystal the 'sphere-like' form of the extreme surface indicates that other crystal cuts can be also used for the realization of the AO cell without essential loss in the diffraction efficiency (about 3% as it is shown by our calculations).

There is the rather unexpected peculiarity, that although the value of $M_2$ increases with decreasing of the acoustic wave velocity, the global maximum of AO figure-of-merit corresponds not only to the slow transversal wave, but it is often achieved on the quasi-longitudinal or the fast quasi-transversal acoustic waves (see Table 1) that characterized by higher velocities.

One of the most interesting results consists in essential increasing (on about 45%) of the AO figure-of-merit for Mg-doped lithium niobate crystal in comparison with the undoped LiNbO$_3$.

## IV. Conclusion

The analysis of the acousto-optic interaction (Bragg diffraction) is carried out for LiNbO$_3$, LiNbO$_3$:MgO, La$_3$Ga$_5$SiO$_{14}$, β-BaB$_2$O$_4$, SrB$_4$O$_7$, Cs$_2$HgCl$_4$, SiO$_2$, CaWO$_4$, Li$_2$B$_4$O$_7$, GaP crystals by extreme surfaces method. At that the optimal geometries of the acousto-optic interaction are determined, i.e. the directions of the electromagnetic incident and acoustic waves that ensuring the highest possible value of the acousto-optic figure-of-merit $M_2$ are defined. Among other considered crystals, the highest value of $M_2$ = 115.9·10$^{-15}$s$^3$/kg is obtained on indirect cut of Cs$_2$HgCl$_4$ crystal.

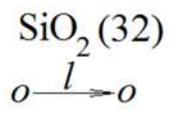
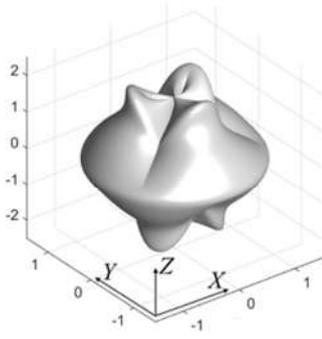
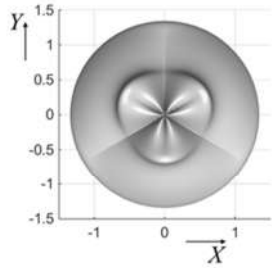
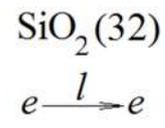
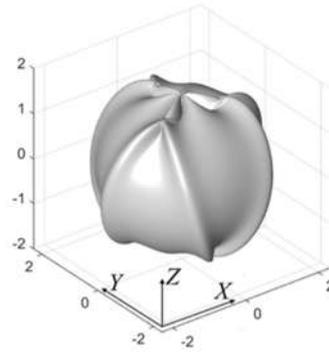
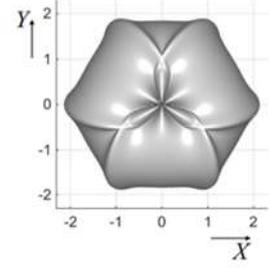
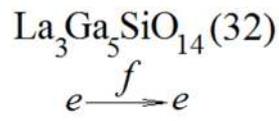
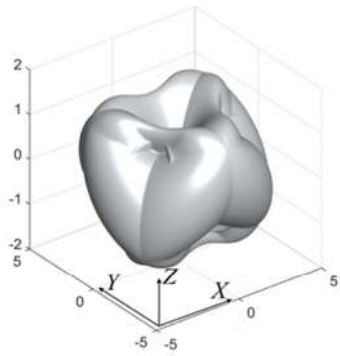
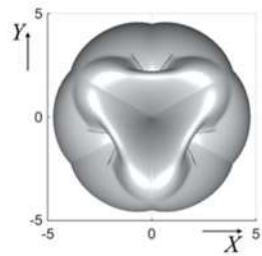
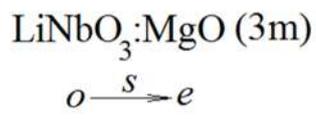
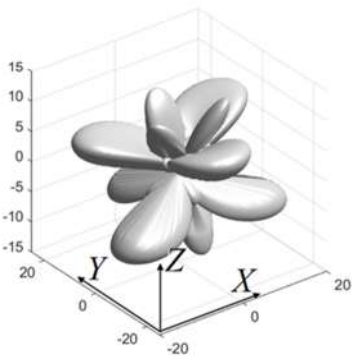
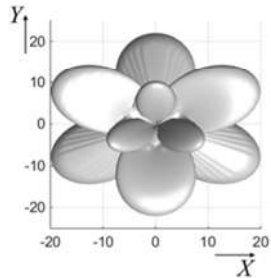
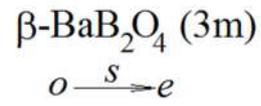
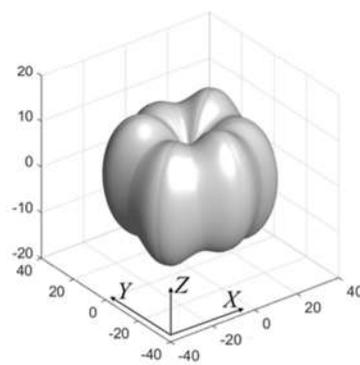
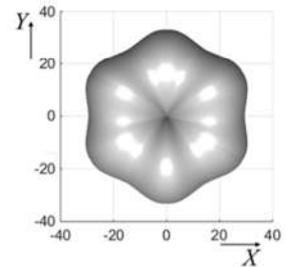

Fig. 1. The AO extreme surfaces for SiO$_2$, La$_3$Ga$_5$SiO$_{14}$, LiNbO$_3$:MgO, β-BaB$_2$O$_4$ trigonal crystals.

## Cs$_2$HgCl$_4$ (mmm)
$i \xrightarrow{l} i$

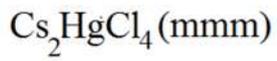
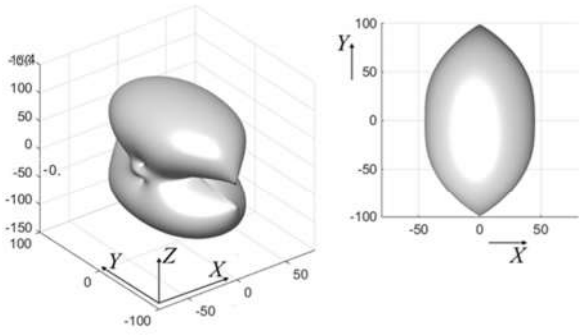

## SrB$_4$O$_7$ (mm2)
$e \xrightarrow{f} e$

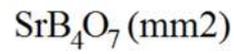
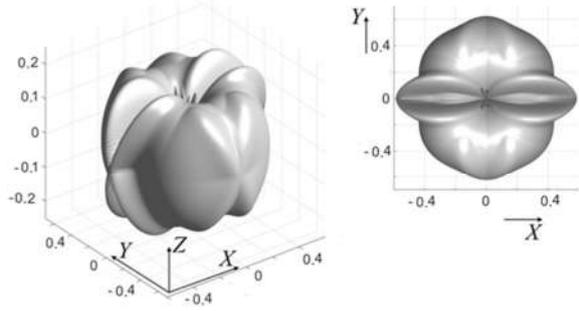

## CaWO$_4$ (4/m)
$o \xrightarrow{s} o$

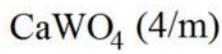
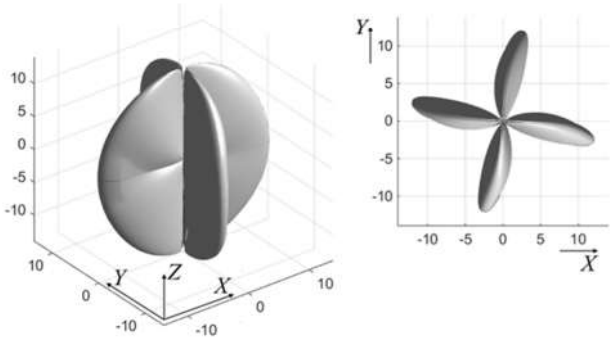

## CaWO$_4$ (4/m)
$e \xrightarrow{s} e$

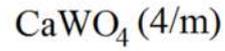
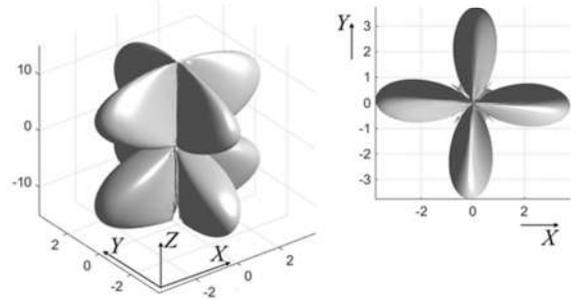

## CaWO$_4$ (4/m)
$o \xrightarrow{s} e$

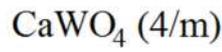
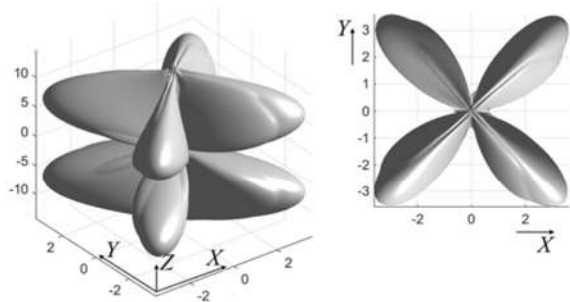

## Li$_2$B$_4$O$_7$ (4mm)
$o \xrightarrow{s} o$

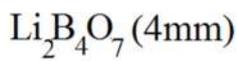
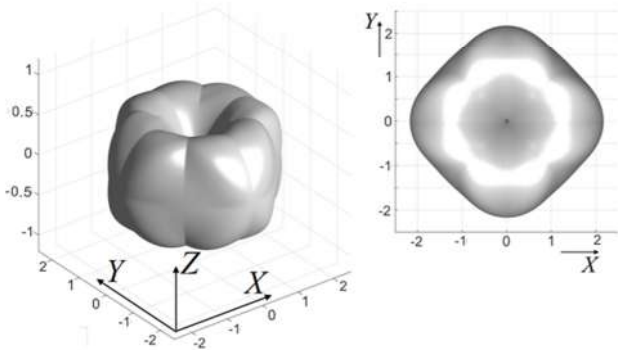

## GaP ($\bar{4}$3m)
isotropic, $l$

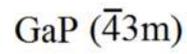
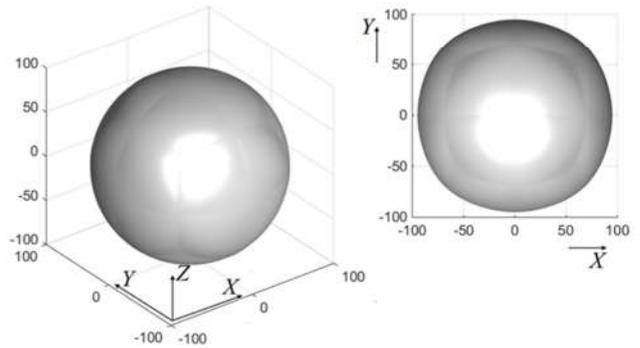

Fig. 2. The AO extreme surfaces for Cs$_2$HgCl$_4$, SrB$_4$O$_7$, CaWO$_4$, Li$_2$B$_4$O$_7$ and GaP crystals.

TABLE I. THE GLOBAL MAXIMA OF THE ACOUSTO-OPTIC EFFECT

| Crystal | Diffraction | $f$, MHz (type of the wave) | Incident light wave | | Acoustic wave | | $V_q$, m/s | $|p_{ef}|$ | $M_{2\,surf}^{max}$, $10^{-15} s^3/kg$ |
|---|---|---|---|---|---|---|---|---|---|
| | | | Direction $\theta_{max}$, $\varphi_{max}$, deg. | Polarization $\theta_{\mu\,max}$, $\varphi_{\mu\,max}$, deg. | Direction $\theta_{a\,max}$, $\varphi_{a\,max}$, deg. | Polarization $\theta_{f\,max}$, $\varphi_{f\,max}$, deg. | | | |
| LiNbO$_3$ | $o \to e$ ($e \to o$) | 500 ($s$) | 73.9; 30 | 90; 120 | 45.1; 30 | 90; 120 | 3546 | 0.16 | 15.9 |
| LiNbO$_3$: MgO | $o \to e$ ($e \to o$) | 500 ($s$) | 73; 30 | 90; 120 | 46.9; 30 | 90; 120 | 3599 | 0.19 | 23.0 |
| La$_3$Ga$_5$SiO$_{14}$ | $e \to e$ | 100 ($f$) | 90; 1 | along OZ | 127.9; 90.6 | 40.4; 90.4 | 3114 | 0.13 | 4.7 |
| β-BaB$_2$O$_4$ | $o \to e$ ($e \to o$) | 100 ($s$) | 87; 27.4 | 90; 117.4 | 90; 120 | 170.1; 29.9 | 882 | 0.08 | 33.2 |
| SrB$_4$O$_7$ | $e \to e$ | 50 ($f$) | along OY | along OZ | 46.9; 179.9 | 130.6; 179.7 | 5999 | 0.14 | 0.63 |
| Cs$_2$HgCl$_4$ | $i \to i$ | 100 ($l$) | 31; 90 | 90; 0 | 120.4; 90 | 119.9; 90 | 1893 | 0.40 | 115.9 |
| SiO$_2$ | $o \to o$ | 100 ($l$) | 18; 30 | 90; 120 | 72.2; 210 | 71.8; 210 | 5322 | 0.25 | 2.1 |
| | $e \to e$ | | 89; 60 | 179; 60 | 106.3; 150 | 99.9; 149.8 | 5331 | 0.25 | |
| CaWO$_4$ | $o \to o$ | 100 ($s$) | 7; 1 | 90; 91 | 91.6; 73.7 | 89.5; 161 | 1883 | 0.11 | 14.0 |
| | $e \to e$ | | 1; 1 | 91; 1 | 91.4; 344 | 89.6; 71 | 1883 | 0.11 | |
| | $o \to e$ ($e \to o$) | | 0.3; 83.1 | 90; 173.1 | 89.6; 163.6 | 89.9; 70.9 | 1882 | 0.11 | |
| Li$_2$B$_4$O$_7$ | $o \to o$ | 100 ($l$) | 90; 89 | 90; 179 | 146.7; -0.1 | 39.2; -0.1 | 3357 | 0.11 | 2.17 |
| GaP | isotropic | 100 ($l$) | 42; 83 | 54.7; 225 | 54.7; 225 | 54.7; 225 | 6521 | 0.30 | 97.6 |

ACKNOWLEDGEMENT

This research has received funding from the European Union's Horizon 2020 research and innovation programme under the Marie Skłodowska-Curie grant agreement No 778156 and was supported by Ministry of Education and Science of Ukraine (projects "Modulator" and "Nanocrystalit", the registration numbers are 0117U004456 and 0119U002255 respectively).